\documentstyle[preprint,aps,floats,epsf,tighten]{revtex}
\begin{document}
\preprint{
\parbox{1.5in}{\leftline{JLAB-THY-99-24}
                \leftline{WM-99-115}
			             \leftline{nucl-th/99xxxxx}
                \leftline{}\leftline{}\leftline{}\leftline{}}}
\title{Charge conjugation invariance
of the Spectator Equations}
\author{Franz Gross}
\address{Department of Physics, College of William and Mary,
 Williamsburg, VA  23185 \\ Thomas Jefferson National Accelerator Facility, 
Newport News, VA 23606}
\date{\today}
\maketitle

\begin{abstract}
In response to recent critcism, we show how to define the spectator
equations for negative energies so that charge conjugation
invariance is preserved.  The result, which emerges naturally from the
application of spectator principles to systems of particles with negative
energies, is to replace all factors of the external energies $W_i$ by
$\sqrt{W^2_i}$, insuring that the amplitudes are independent of the sign of the
energies $W_i$.  
\end{abstract}

\widetext

\section{Introduction}

In a recent set of papers\cite{PT1,PT2}, Pascalutsa and Tjon have criticized the
spectator formalism by claiming that it violates charge conjugation
invariance, C.  When applied to the self energy of a Dirac particle,
$\Sigma(p_0,{\bf p})$,  this requirement is
\begin{eqnarray}
C\Sigma(p_0,{\bf p})C^{-1}=\Sigma^T(-p_0,-{\bf p})\, ,
\label{1eq1}
\end{eqnarray} 
where $C$ is the Dirac charge conjugation matrix and the superscript $T$
refers to the transpose in the Dirac space.  The spectator equations have
been previously applied only to the positive energy subspace, and the
transformation (\ref{1eq1}) is the only one of all the transformations
in the full Lorentz group that connects states of positive
and negative energy.  Before it can be tested, the definition
of the spectator equations must be extended to negative energy.  The
claim of Pascalutsa and Tjon that the spectator formalism violates charge
conjugation invariance (and hence Lorentz covariance) follows from
their consideration of how the spectator equations should be extended to
negative energy.

In this short paper we confirm that the extension of the spectator
formalism to negative energies proposed by Pascalutsa and Tjon does indeed
violate C invariance, but that a more natural extension does not.  Since the
equations have never been applied to negative energies before the work of
Refs.~\cite{PT1,PT2}, our discussion is, strictly speaking, a proposal for how
the equations should be extended to negative energies in such a way as to
preserve C invariance.  We will show that this extension is a natural and
a fathful application of the basic principles guiding the construction of the
spectator theory.

In the next section we review the basic principles underlying the
spectator theory \cite{gross}, and apply these principles to the study of
systems with negative energy.  This leads naturally to the principle that
all external energies, referred to collectively as $W_i$, should be
interpreted as $|W_i|$ (or $\sqrt{W_i^2}$), insuring that the C invariance
condition (\ref{1eq1}) is trivially satisfied.  In Sec.~III we present some
examples in 1+1 dimension, where numerical results can be easily obtained
without the use of form factors.  We summarize our conclusions in a final
section.

\section{Principles of the Spectator Theory and its Extension to Negative
Energies}

\begin{figure}
\centerline
{\epsfxsize=4in\epsffile{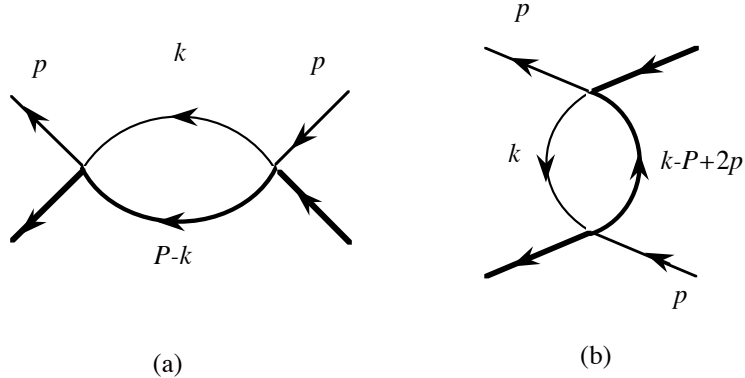}}
\caption{The bubble (a) and ``crossed
bubble'' (b) diagrams.}\label{bubbles}
\end{figure}

The principles of the spectator theory are illustrated by a simple
$\lambda\,\phi^2\Psi^\dagger\Psi$ theory, where $\Psi$ is a ``heavy'' fermion
field of mass $m$ (refered to as a ``nucleon'') and $\phi$ is a ``light''
self-conjugate boson with mass $\mu$ (refered to as a ``meson'').  The one-loop
diagrams in the theory are the second order bubble diagrams shown in
Fig.~\ref{bubbles}.  Here $P=(W, {\bf 0})$ is the total four-momentum of the
pair, and $p$ is the external four-momentum of the heavy Dirac particle. 
While we assume a scalar interaction for simplicity, all of our results are
independent of the Dirc structure of the interaction. 

The Feynman integral for the bubble diagram (a) is
\begin{eqnarray}
\Sigma_a(P)=i\lambda^2\int \frac{d^4k}{(2\pi)^4} \frac{(m+\backslash \!\!\!
P-\backslash \!\!\! k)}  {(\mu^2-k^2)(m^2-(P-k)^2)} F(k^2, k\cdot P, P^2) \, ,
\label{2eq3}
\end{eqnarray} 
where $F$ is a function that depends on the form factors or 
regularization prescription used in the calculation, and must depend on the
arguments $k^2, k\cdot P,$ and $P^2$.  Changing $P\to -P$ and $k\to -k$ 
shows immediately that 
\begin{eqnarray}
\Sigma_a(-P) = C\Sigma_a^T(P)C^{-1}\, ,
\label{2eq4}
\end{eqnarray} 
and the diagram is invariant under charge conjugation.  A similar
argument works for the crossed bubble (b).  Hence the four-dimensional
calculation of these diagrams is C invariant.

Now look at the spectator calculation of these two diagrams.  The philosophy
underlying the spectator approach, as {\it commonly stated\/}, is to
approximate the diagram (a) by picking up the leading positive energy heavy
particle pole, and to lump all other contributions from these diagrams with
the higher order terms included if the calculation were to be carried out to
third (or higher) order.  Of course, if we only need the result to second
order (for example, when calculating  high energy scattering for a weak
coupling when perturbation theory gives a reliable result) it is simple enough
to obtain the exact answer in this case.  But in the more general case (for
example, when the coupling is strong or an infinite sum of diagrams is needed
at low energy of near bound state poles -- even when the coupling is small)
then we will need a systematic approach which sums all ladder and crossed
ladder diagrams efficiently.  For this simple theory, the bubble diagram (a)
plays the role of a fourth order ladder diagram [where 
\begin{eqnarray}
\lambda = -\frac{g^2}{M^2-q^2}\to -\frac{g^2}{M^2}
\label{2eq5}
\end{eqnarray} 
is the effective coupling from a very heavy meson exchange of very short
range] and the crossed bubble plays the role of the fourth order crossed
ladder.  [This can be easily demonstrated by writing down these diagrams and
letting the heavy meson mass, $M\to\infty$].

\begin{figure}
\centerline
{\epsfxsize=4in\epsffile{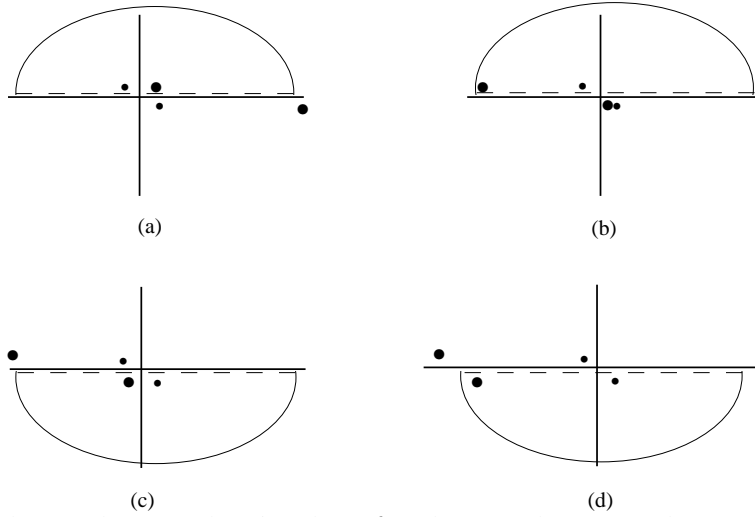}}
\caption{The poles in the complex $k_0$ plane for the two
diagrams shown in Fig.~1.  Panel (a) shows the poles for diagram 1(a) and
panel (b) the poles for diagram 1(b) when $W=|W|>0$.  Panels (c) and (d) 
show the corresponding locations of the poles when
$W=-|W|<0$.}\label{poles}
\end{figure}

Understanding of the mathematical and physical motivation behind the spectator
theory comes from a study of the singularities of the two bubble diagrams in
the complex $k_0$ plane.  In the next section we will give a numerical 
demonstration of the following discussion; here we focus on a qualitative
understanding.   The two bubbles each have four poles in
the complex $k_0$ plane.  In the rest  frame of the two particles, the four
poles for diagram \ref{bubbles}(a) are at
\begin{eqnarray}
k_0&=&\pm\left(\omega(k)-i\epsilon\right) 
\nonumber\\  
&=&W-E(k)+i\epsilon 
\nonumber\\ 
&=&W+E(k)-i\epsilon 
\, ,
\label{2eq1}
\end{eqnarray} 
where $\omega(k)=\sqrt{\mu^2+{\bf k}^2}$ and $E(k)=\sqrt{m^2+{\bf k}^2}$. 
For diagram \ref{bubbles}(b), the poles are at
\begin{eqnarray}
k_0&=&\pm\left(\omega(k)-i\epsilon\right) 
\nonumber\\ 
&=&W-2E(p)-E(k)+i\epsilon 
\nonumber\\
&=&W-2E(p)+E(k)-i\epsilon 
\, .
\label{2eq2}
\end{eqnarray} 
The location of these poles is shown in Fig.~\ref{poles} for the case when
${\bf k}\simeq0$, ${\bf p}\simeq0$, and $|W|\simeq m+\mu$.  When $W>0$
[panels (a) and (b)] the coutour shown in the figures is closed in the
upper half plane, and encloses the negative energy
meson pole and [in panel (a)] the positive energy nucleon pole or [in panel
(b)] the negative energy nucleon pole.  In panels (c) and (d) the opposite is
true; the contour is closed in the lower half plane and encloses the positive
energy
meson pole and [in panel (c)] the negative energy nucleon pole or [in panel
(d)] the positive energy nucleon pole.

First suppose that the external energy is positive, and $W\simeq m+\mu$, so
that the location of the singularities is as given in panels (a) and (b). 
Closing the contour for diagram (a) in the upper half plane gives two
contributions: the positive energy nucleon pole and the negative energy meson
pole.  At first it looks like the negative energy meson pole will introduce a
large correction (because it is so close to the nucleon pole), but it
turns out that {\it the contribution from the negative energy meson pole is
almost exactly cancelled by a similar contribution from the crossed bubble
(b), and hence the nucleon pole alone gives a very accurate result\/}.  This
will be demonstrated numerically in the next section.  Hence, for $W>0$, 
the spectator result for {\it both\/} of the bubble diagrams in
Fig.~\ref{bubbles} is the nucleon pole contribution from (a) 
\begin{eqnarray}
\Sigma_S(W)\Big|_{W>0}=-\lambda^2\int \frac{d^3k}{(2\pi)^3}
\frac{(m+\backslash
\!\!\! P-\backslash \!\!\! \hat{k})} 
{2E(k)[\omega^2(k)-\left(W-E(k)\right)^2]} F(\hat{k}^2, \hat{k}\cdot P, P^2)
\, ,
\label{2eq6}
\end{eqnarray} 
where $\hat{k}=(W-E(k),-{\bf k})$.  Since the internal
energy of the nucleon is positive, the internal energy cannot be changed and
the argument used to demonstrate C invariance for the four-dimensional
calculation fails.

However, {\it the same physics which lead to the selection of the positive
energy nucleon pole will yield a different result\/} if the
bubbles are to be evaluated at a negative external energy $W\simeq
-(m+\mu)$.  Since negative external energies are unphysical, this case was
not considered in the original formulation of the spectator theory.  We
look at it now.

For negative external energies $W\simeq-(m+\mu)$ the poles are as shown in
panels (c) and (d).  Now the role of the upper and lower half planes are
changed, and the {\it negative energy\/} nucleon pole dominates diagram (a),
with the {\it positive\/} energy meson pole giving the leading correction. 
Furthermore, as in the positive energy case, this leading correction (from the
positive energy meson pole now) is cancelled by the contribution from the
crossed bubble.  Hence, for both diagrams {\it the same
physical/mathematical argument\/} yields, for negative energy,
\begin{eqnarray}
\Sigma_S(W)\Big|_{W<0}=-\lambda^2\int \frac{d^3k}{(2\pi)^3}
\frac{(m+\backslash
\!\!\! P-\backslash \!\!\! \tilde{k})} 
{2E(k)[\omega^2(k)-\left(W+E(k)\right)^2]} F(\tilde{k}^2, \tilde{k}\cdot P,
P^2)
\, ,
\label{2eq7}
\end{eqnarray} 
where $\tilde{k}=(W+E(k),-{\bf k})$.  Comparing Eqs.~(\ref{2eq6}) and
(\ref{2eq7}) [after changing ${\bf k}\to -{\bf k}$] shows that
\begin{eqnarray}
\Sigma_S(W)\Big|_{W<0}=\Sigma_S(-|W|)\Big|_{W<0} =
C\Sigma_S^T(W)\Big|_{W>0}C^{-1}\, .
\label{2eq8}
\end{eqnarray} 
This is the proof of C invariance we seek.  Note that the natural extension
of the spectator equations to negative energy has lead to a result
which can be obtained from the positive energy result by using the
transformation (\ref{2eq8}).   

If we separate the $\Sigma_S$ into scalar functions according to
\begin{eqnarray}
&&\Sigma_S(W)\Big|_{W>0}=\backslash\!\!\! P 
A(W)\Big|_{W>0}+B(W)\Big|_{W>0}\nonumber\\
&&\Sigma_S(W)\Big|_{W<0} = \backslash\!\!\! P 
A(W)\Big|_{W<0}+B(W)\Big|_{W<0}\, ,
\label{2eq9}
\end{eqnarray} 
then, from Eqs.~(\ref{2eq6}) and (\ref{2eq7})
\begin{eqnarray}
B(W)\Big|_{W>0}&=&-\lambda^2\int \frac{d^3k}{(2\pi)^3} \frac{m\;
F(\hat{k}^2, \hat{k}\cdot P,  P^2)}
{2E(k)[\omega^2(k)-\left(W-E(k)\right)^2]} \nonumber\\
B(W)\Big|_{W<0}&=&-\lambda^2\int \frac{d^3k}{(2\pi)^3} \frac{m\;
F(\tilde{k}^2,\tilde{k}\cdot P,  P^2)}
{2E(k)[\omega^2(k)-\left(W+E(k)\right)^2]}
\nonumber\\ 
A(W)\Big|_{W>0}&=&-\lambda^2\int \frac{d^3k}{(2\pi)^3}
\frac{F(\hat{k}^2,\hat{k}\cdot P,  P^2)}
{2W[\omega^2(k)-\left(W-E(k)\right)^2]} \nonumber\\ 
A(W)\Big|_{W<0}&=&+\lambda^2\int \frac{d^3k}{(2\pi)^3} \frac{
F(\tilde{k}^2,\tilde{k}\cdot P,  P^2)}
{2W[\omega^2(k)-\left(W+E(k)\right)^2]} \, .
\label{2eq10}
\end{eqnarray} 
Hence
\begin{eqnarray}
&&B(W)\Big|_{W<0}=B(W)\Big|_{W>0}\nonumber\\
&&A(W)\Big|_{W<0}=A(W)\Big|_{W>0} \, .
\label{2eq11}
\end{eqnarray} 
These are precisely the properties of the scalar functions $A$ and $B$
required by C invariance.  They are possible because $A$ and $B$ for $W<0$
are {\it different algebraic functions\/} of $W$.  The simple
relationships (\ref{2eq11}) between the functions for
$W<0$ and $W>0$ permits us to write them as a single function of $|W|$: 
\begin{eqnarray}
&&B(W)\Big|_{W<0}=B(W)\Big|_{W>0}\equiv B(|W|)=B(\sqrt{W^2})\nonumber\\
&&A(W)\Big|_{W<0}=A(W)\Big|_{W>0}\equiv A(|W|)=A(\sqrt{W^2}) \, ,
\label{2eq12}
\end{eqnarray} 
as stated in the introduction.  While the rule (\ref{2eq12}) was only
derived in this section for a simple $\varphi^4$-type theory,
examination of the details of the derivation will convince one that it
can be extended to the general case.

We turn now to short numerical study of these results.

\section{Numerical Examples in 1+1 Dimension}

The discussion in the last section showed that the natural extension of
the spectator equations to negative energies preserves charge conjugation
invariance.  In this section show that

\begin{itemize}
\item failure to use the prescription $W\to|W|$ when applying the
spectator theory to negative energies does indeed lead to very serious
numerical errors, as pointed out in Refs.~\cite{PT1,PT2}, and

\item the spectator approximation to the {\it sum\/} of the bubble and
the crossed bubble is a better approximation that the exact bubble diagram
itself.

\end{itemize}

In order to keep the discussion simple and to the point, we limit these
numerical examples to the $B$ function in 1+1 dimensions, where
the integrals converge without form factors \cite{ZFBG1}.  Extension of the
results to the $A$ function, and to higher dimensions yields similar
results, but is complicated by the need for form factors or cutoffs. 

\begin{figure}
\centerline
{\epsfxsize=4in\epsffile{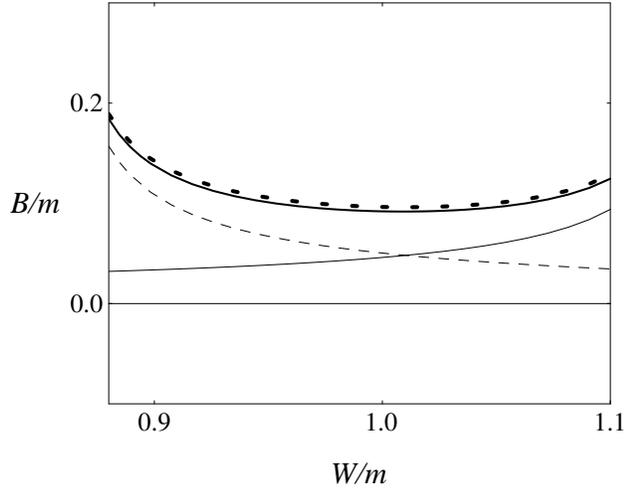}}
\caption{The scalar functions $B/m$ for the bubble and ``crossed
bubble'' shown in Fig.~1.  The light solid line is the exact bubble, $B_a$,
the dashed line is the crossed bubble, $B_b$ (for forward scattering), the
heavy dotted line is the sum,  and the  heavy solid line is the positive
energy nucleon pole contribution, $B_S$.}\label{three}
\end{figure}

In 1+1 dimension, the $B$ functions for diagrams \ref{bubbles}(a) and (b)
are
\begin{eqnarray}
&&B_a(P)=i\lambda^2\int \frac{d^2k}{(2\pi)^2} \frac{m} 
{(\mu^2-k^2)(m^2-(P-k)^2)} \nonumber\\
&&B_b(P)=i\lambda^2\int \frac{d^2k}{(2\pi)^2} \frac{m} 
{(\mu^2-k^2)(m^2-(P-2p-k)^2)} \, ,
\label{3eq1}
\end{eqnarray} 
where the form factor function has been set to unity.  These integrals are
easily evaluated.  Numerical results for the case when $M=m/\mu=7$,
$\lambda^2/(2\pi\mu^2)=3$, $p=(E(p), p)$, and $W=E(p)+\omega(p)$,
corresponding to scattering in the forward direction, are shown in in
Fig.~\ref{three}.  Note that the bubble (a) and crossed bubble (b) are
comparable in size, and that their sum (the heavy dotted line) is almost
identical to the positive energy (because $W>0$!) nucleon pole contribution
from diagram (a) alone.  This latter is 
\begin{eqnarray}
B_S(W)&=&-\frac{m\lambda^2}{4\pi}\int \frac{dk} 
{E(k)[\mu^2-m^2 -W^2 +2E(k)W]} \nonumber\\
&=&-\frac{m\lambda^2}{2\pi\mu^2}\int_0^\infty \frac{d\kappa} 
{e(\kappa)[1-M^2 -(W/\mu)^2 +2e(\kappa)(W/\mu)]} \, ,
\label{3eq3}
\end{eqnarray} 
where $k=\mu\kappa$ and $e(\kappa)=\sqrt{M^2+\kappa^2}$ is the 
dimensionless form of $E(k)$.  

The Figure shows clearly that the spectator contribution gives a much better
description of the sum of the direct and crossed bubbles than that given by
the direct bubble alone.  The reason is that the (large) contribution from
the nearby negative energy meson pole is cancelled by a similar
contribution from the crossed bubble diagram.  Such a cancellation
between ladder and crosses ladder diagrams is the foundation of the
spectator theory \cite{gross}, but is was initially assumed that this
cancellation would only be important when light mesons were exchanged
between the scattered particles.  Since a four point interaction is
equivalent to the exchange of an infinitely heavy meson [cf.
Eq.~(\ref{2eq5}) above], this discussion demonstrates that the same
cancellation is also important {\it even if the effective mesons being
exchanged are very heavy\/}.  This has been recently noted by
Pascalutsa and Tjon \cite{PT2} and by the author \cite{JLablec}.

What should we do it there is no exchange term (exchange bubble in our
example)?  This situation could arise in a $\varphi^3$-type theory, for
example the familiar theory $\Psi^\dagger\gamma_5\Psi\phi$.  In this case
the lowest order self energy diagram would be a bubble of the type shown
in Fig.~\ref{bubbles}(a), and there would be no contribution similar to
diagram \ref{bubbles}(b).  Hence there is {\it no diagram to cancel
the negative energy meson pole\/}, and the taking the positive energy
nucleon pole will not give a good description of the full result.  For this
reason, Surya and I \cite{SG1,SG2} decided to use spectator equations
based on the positive energy {\it meson\/} pole, which would emerge by
closing the contour in Fig.~\ref{poles}(a) in the {\it lower\/} half
plane.  The positive energy meson pole which is isolated in this way is
very distant from the negative energy nucleon pole and gives a good
approximation to the exact bubble. Of course one could just
asd well calculate the bubble exactly, but we intended to eventually imbed 
the equations in the $NN\pi$ system, and wanted to preserve the spectator
formalism in a three body system \cite{SG}.

\begin{figure}
\centerline
{\epsfxsize=4in\epsffile{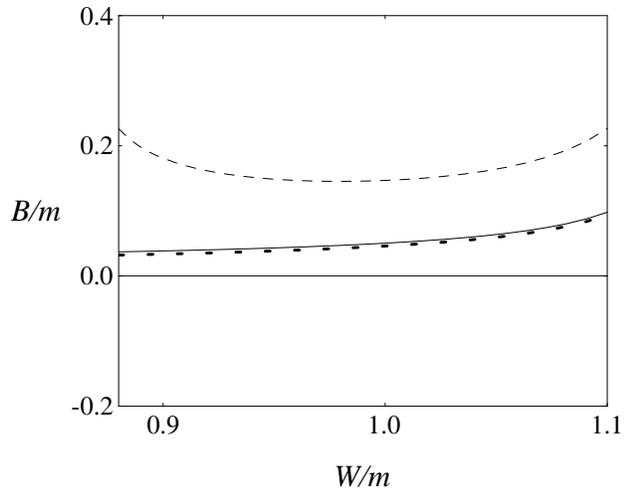}}
\caption{The scalar function $B/m$ as a function of the scaled
energy $W/m$.  The solid line is the positive energy meson pole 
contribution to $B_a(W)$; the dashed line is the {\it negative energy\/}
meson pole contribution, and the dotted line is the exact result
(identical to Fig.~3).}\label{one}
\end{figure}

As shown in Fig.~\ref{one}, the positive energy meson pole does an
excellent job approximating the exact result for the direct bubble
\ref{bubbles}(a) [the solid line and the dotted line agree very well]. 
However, if $W<0$, the positive energy meson pole gives a very
different result.  This {\it positive\/} energy meson pole contribution for
negative $W$ is {\it identical\/} to the {\it negative\/} energy meson pole
contribution for {\it positive\/} $W$, and this is the result shown (the
dashed line) in the figure.  We see, in agreement with
Refs.~\cite{PT1,PT2}, that using the positive energy pole for {\it both\/}
$W>0$ and $W<0$ violates C invarinace very significantly.  However, if the
positive energy pole is used when $W>0$ and the negative energy pole when
$W<0$, as suggested by the spectator philosophy, the results are identical
and C invarinace is satisfied.  This brings us back to the discussion and
derivation in the previous section.

\section{Conclusions}

In this paper we have shown that 

\begin{itemize}

\item  the spectator equations satisfy charge conjugation
invariance exactly provided the on-shell internal particle(s) are on their
positive energy mass-shell when the external energies are positive, and
on their negative energy mass-shell when the external energies are
negative;

\item this requirement is equivalent to extending the positive energy
spectator theory to negative energies by replacing all external energies
$W_i$ by $|W_i|=\sqrt{W_i^2}$, and is consistent with the spectator
philosophy;

\item spectator equations with the the heavy particle on-shell should
be used in all cases when there are exchange forces; and

\item spectator equations with the light particle on-shell (or the
Bethe-Salpeter equation) should be used if there are no exchange forces. 

\end{itemize}

We close this discussion by emphasizing that the simplified {\it one channel\/}
spectator theory described in this paper cannot be used when it is important to
get an accurate description of the self energies or scattering amplitudes in a
region where the external energy $W$ is near zero.  The simplified
treatment described here has unphysical singularities at $W=0$ [clearly
evident in Eq.~(\ref{2eq7})], and unphysical cuts for $W^2<0$. In
studies of the pion, where chiral symmetry requires an accurate description near
$m_\pi=W\simeq0$, and C invariance requires that the $q\bar{q}$ system be
treated symmetrically, the {\it four channel\/} spectator equation originally
introduced in Ref.~\cite{GM} must be used.

\acknowledgments

This work is submitted to this issue of {\it Few Body Systems\/} in
celebration of Walter Gl\"ockle's  60th birthday, and in recognition of
the important contributions he has made to the study of relativistic
theories of few body systems.  It is a pleasure to acknowledge the many
valuable discussions I have had with Walter over the years.

This research has been supported in part by the DOE under grant
No.~DE-FG02-97ER41032, and by DOE contract DE-AC05-84ER40150  administered
by SURA in support of the Thomas Jefferson National Accelerator
Facility.  I also wish to thank Fritz Co\"ester for sharing his many 
insights with me, and V.~Pascalutsa and John Tjon for helpful discussions.
 



\end{document}